\documentclass[12pt]{article}
\usepackage{amsmath,amssymb,bm,graphicx}
\begin{document}

\begin{flushright}
\end{flushright}

\vskip 0.5 truecm

\begin{center}
{\Large{\bf Does CHSH inequality test the model of 
local hidden variables
?}}
\end{center}
\vskip .5 truecm
\centerline{\bf  Kazuo Fujikawa}
\vskip .4 truecm
\centerline {\it Institute of Quantum Science, College of 
Science and Technology}
\centerline {\it Nihon University, Chiyoda-ku, Tokyo 101-8308, 
Japan\footnote{Address after April 1, 2012: Mathematical Physics Laboratory, RIKEN Nishina Center, Wako, Saitama 351-0198, Japan}}
\vskip 0.5 truecm

\makeatletter
\@addtoreset{equation}{section}
\def\theequation{\thesection.\arabic{equation}}
\makeatother

\begin{abstract}
It is pointed out that the local hidden variables model of Bell and Clauser-Horne-Shimony-Holt (CHSH) gives $|\langle B\rangle |\leq 2\sqrt{2}$ or $|\langle B\rangle |\leq 2$ for the quantum CHSH operator $B={\bf a}\cdot {\bf \sigma}\otimes ({\bf b}+{\bf b}^{\prime})\cdot {\bf \sigma} +{\bf a}^{\prime}\cdot{\bf \sigma}\otimes ({\bf b}-{\bf b}^{\prime})\cdot{\bf \sigma}
$ depending on two different ways of evaluation, when it is applied to a  $d=4$ system of two spin-$1/2$ particles. This is due to the failure of linearity, and it shows that  the conventional CHSH inequality $|\langle B\rangle |\leq 2$ does not provide a reliable test of the $d=4$ local  non-contextual hidden variables model. To achieve $|\langle B\rangle |\leq 2$ uniquely, one needs to impose a linearity requirement  
on the hidden variables model, which in turn adds a von Neumann-type stricture. It is then shown that the local model is converted to a factored  product of two non-contextual $d=2$ hidden variables models. This factored product implies pure separable  quantum states and satisfies $|\langle B\rangle |\leq 2$, but no more a proper hidden variables model in $d=4$. The conventional CHSH inequality $|\langle B\rangle |\leq 2$ 
thus characterizes the pure separable quantum mechanical states but 
does not test the model of local hidden variables in $d=4$, to be consistent with Gleason's theorem which excludes non-contextual models in $d=4$. This observation is also consistent with an application of the CHSH inequality to quantum cryptography by Ekert, which is based on mixed separable states without referring to hidden variables.  
\end{abstract}

\section{Introduction}
It is generally considered that Bell~\cite{bell2} and Clauser-Horne-Shimony-Holt (CHSH)~\cite{chsh} inequalities, which were introduced to analyze EPR entanglement~\cite{epr}, compare the prediction of quantum mechanics with the prediction of any theory which is based on local realism. For example, one usually starts with the relation for dichotomic variables~\cite{peres} $a_{j}(b_{j}+b^{\prime}_{j})+a^{\prime}_{j}(b_{j}-b^{\prime}_{j})=\pm 2$, and one sums the dichotomic variables with a {\em uniform} weight factor $P_{j}\geq 0$ for all the terms with $\sum_{j}P_{j}=1$ to obtain CHSH inequality
\begin{eqnarray}
|\langle ab\rangle+\langle ab^{\prime}\rangle+\langle a^{\prime}b\rangle-\langle a^{\prime}b^{\prime}\rangle|\leq 2
\end{eqnarray}
with $\langle ab\rangle=\sum_{j}P_{j}a_{j}b_{j}$.
The uniform weight for all the combinations of dichotomic variables manifests the strict locality which also implies  non-contextuality~\cite{gleason, bell1, kochen}. This formulation however does not test local hidden variables models in quantum mechanics in general. Hidden variables models in quantum mechanics mean that they satisfy certain basic properties of quantum mechanics which are not incorporated in the above dichotomic variables. In $d=2$, where $d$ stands for the dimensionality of the Hilbert space,  we have rather satisfactory hidden variables models which simulate quantum mechanics~\cite{bell1,kochen}, but we have no definite explicit hidden variables models which describe the essence of quantum mechanics in $d=4$. Bell and CHSH thus consider a generic local and non-contextual hidden variables model in $d=4$, which is supposed to describe the local properties of quantum mechanics but not necessarily non-local properties such as entanglement~\cite{bell2,chsh}. One can then test entanglement, which is the characteristic feature of quantum mechanics, by comparing the predictions of quantum mechanics and the hidden variables model with experiments~\cite{ch}.     
 
In the analysis of the hidden variables model, we use for definiteness the quantum CHSH operator~\cite{cirel'son}
\begin{eqnarray}
B={\bf a}\cdot {\bf \sigma}\otimes ({\bf b}+{\bf b}^{\prime})\cdot {\bf \sigma} +{\bf a}^{\prime}\cdot{\bf \sigma}\otimes ({\bf b}-{\bf b}^{\prime})\cdot{\bf \sigma}\nonumber
\end{eqnarray}
defined for a system of two spin-$1/2$ particles by considering the system as a $d=4$ dimensional one. It is then pointed out that the local non-contextual hidden variables model of Bell and CHSH gives rise to $|\langle B\rangle|\leq 2\sqrt{2}$ or  $|\langle B\rangle|\leq 2$ for the identical CHSH operator depending on two different ways of evaluation, when it is applied to this $d=4$ system.  This is due to the lack of linearity, and it shows that the conventional  CHSH inequality $|\langle B\rangle |\leq 2$ does not provide a reliable test of the $d=4$ local non-contextual hidden variables model in quantum mechanics. To achieve $|\langle B\rangle|\leq 2$ uniquely, one needs to impose a linearity requirement of quantum mechanics on the model, which in turn adds a von Neumann-type stricture. 
The non-contextual model in $d=4$ is then inevitably converted to the contextual one with extra constraints on the weight factor in the hidden variables model. The non-contextual model of Bell and CHSH is converted to a factored  product of two non-contextual $d=2$ hidden variables models, which implies pure quantum mechanical separable states, to be consistent with Gleason's theorem~\cite{gleason} excluding fully $d=4$ non-contextual models. 

This  analysis of separable states is related to  the well-known analysis of Werner~\cite{werner}. By adopting the definition of local hidden variables model of Bell and CHSH in $d=4$, Werner shows that the conventional CHSH inequality provides the necessary and sufficient separability criterion for pure quantum mechanical states. Moreover, he shows that a certain mixed inseparable state satisfies the conventional CHSH inequality.     

\section{Hidden variables model}
We start with a brief review of hidden variables models.
Bell's explicit construction in $d=2$ illustrates what a hidden variables model is in a concrete manner~\cite{bell1}. It is based on the projector 
\begin{eqnarray}
P_{{\bf m}}=\frac{1}{2}(1+{\bf m}\cdot{\bf \sigma})
\end{eqnarray}
with a unit vector $|{\bf m}|=1$ and Pauli matrix ${\bf \sigma}$, and the  dispersion free representation of $P_{{\bf m}}$ which assumes eigenvalues $1$ or $0$ depending on the hidden parameter ${\bf \omega}$ in the interval $\frac{1}{2}\geq \omega\geq -\frac{1}{2}$ as,
\begin{eqnarray}
P_{{\bf m} \psi}(\omega)=\frac{1}{2}[1+\text{sign}(\omega+\frac{1}{2}|{\bf s}\cdot {\bf m}|)\text{sign}({\bf s}\cdot {\bf m})]
\end{eqnarray}
for the pure state represented by the projector $|\psi\rangle\langle\psi|=\frac{1}{2}(1+{\bf s}\cdot{\bf \sigma})$ with $|{\bf s}|=1$. This $P_{{\bf m} \psi}(\omega)$ reproduces the quantum mechanical result after integration over $\omega$ (i.e., a uniform {\em non-contextual} weight $\rho(\omega)=1$ for the hidden variable $\omega$ in the present example)
\begin{eqnarray}
\int_{-1/2}^{1/2} P_{{\bf m} \psi}(\omega)d\omega =\langle\psi|P_{{\bf m}}|\psi\rangle.
\end{eqnarray}
 It is shown that the dispersion free $P_{{\bf m} \psi}(\omega)$ itself is not given by any density matrix parameterized by ${\bf s}$ and $\omega$~\cite{beltrametti}. For a general $2\times 2$ hermitian operator $O$
 which can be written in a spectral decomposition
$O=\mu_{1}P_{1}+\mu_{2}P_{2}$
with two orthogonal projectors $P_{1}$ and $P_{2}$, $P_{1}+P_{2}=1$, the dispersion free representation is consistently defined by
\begin{eqnarray}
O_{\psi}(\omega)=\mu_{1}P_{1,\psi}(\omega)+\mu_{2}P_{2,\psi}(\omega).
\end{eqnarray}
Note that $P_{1,\psi}(\omega)+P_{2,\psi}(\omega)=1$.
The essence of Bell's hidden variables model is that it gives a dispersion free representation of a general hermitian operator and that it reproduces the quantum mechanical result for any pure single-particle states after integrating over the hidden variable. 
It is considered that Bell's explicit construction of the non-contextual model in $d=2$ is free from the existing no-go theorems~\cite{gleason, kochen, beltrametti, peres}. See, however, the recent analysis of this issue~\cite{fujikawa}.

One may next consider a linear combination of two non-collinear projectors in (2.1)
\begin{eqnarray}
E=\lambda P_{{\bf n}}+ (1-\lambda) P_{{\bf m}}, \ \ \ \  0< \lambda< 1,
\end{eqnarray}
which satisfies $0<E<1$. If one assumes that the dispersion free representation of Bell is applied to all the operators in (2.5) separately, one obtains
\begin{eqnarray}
E_{\psi}(\omega)=\lambda P_{{\bf n} \psi}(\omega)+ (1-\lambda) P_{{\bf m} \psi}(\omega), \ \ \ \ 0< \lambda< 1,
\end{eqnarray}
but this is not satisfied by the positive operator  $1>E_{\psi}(\omega)>0$ on the left-hand side in the domain of the hidden variables space
with $P_{{\bf n} \psi}(\omega)=P_{{\bf m} \psi}(\omega)=0$ (or with
$P_{{\bf n} \psi}(\omega)=P_{{\bf m} \psi}(\omega)=1$). This shows that Bell's construction cannot maintain the linearity (2.6) at each point of hidden variables space, although it reproduces the result of quantum mechanics 
\begin{eqnarray}
\langle E\rangle_{\psi}=\lambda\langle  P_{{\bf n}}\rangle_{\psi}+ (1-\lambda) \langle P_{{\bf m}}\rangle_{\psi}
\end{eqnarray}
implied by (2.5) after integrating over hidden variables.

The conflict in (2.6) is essentially the original no-go argument of von Neumann~\cite{neumann} against non-contextual hidden variables models, and its physical resolution is well known. One does not assign a physical significance to the dispersion free representation of each operator such as in (2.6) simultaneously in hidden variables space, since two non-commuting operators are not physically compatible~\cite{bell1}. The stricture of von Neumann-type  plays a central role in our analysis although we assign a physical significance only to  integrated quantities. 

We next comment on a system of two spin-$1/2$ particles in the analysis of CHSH inequality. 
For a $d=4$ system under certain assumptions, 
we have Gleason's theorem~\cite{gleason} which states that the quantum probability is defined by a trace representation with a suitable density matrix ${\text Tr}\rho O$  for any hermitian operator $O$,
and that non-contextual hidden variables models do not exist in general~\cite{bell1, kochen, beltrametti, mermin}.

\section{ CHSH operator}

To make the framework of our analysis definite, we work with the CHSH operator $B$ introduced by Cirel'son~\cite{cirel'son}
\begin{eqnarray}
B={\bf a}\cdot {\bf \sigma}\otimes ({\bf b}+{\bf b}^{\prime})\cdot {\bf \sigma} +{\bf a}^{\prime}\cdot{\bf \sigma}\otimes ({\bf b}-{\bf b}^{\prime})\cdot{\bf \sigma}
\end{eqnarray}
with 3-dimensional unit vectors ${\bf a},\ {\bf a}^{\prime},\ {\bf b},\ {\bf b}^{\prime}$, and ${\bf \sigma}$ standing for the Pauli matrix.
 One can first establish
\begin{eqnarray}
||B||\leq 2\sqrt{2}
\end{eqnarray}
by noting 
\begin{eqnarray}
&&||{\bf a}\cdot {\bf \sigma}\otimes ({\bf b}+{\bf b}^{\prime})\cdot {\bf \sigma}||\leq |{\bf b}+{\bf b}^{\prime}|,\nonumber\\
&&||{\bf a}^{\prime}\cdot{\bf \sigma}\otimes ({\bf b}-{\bf b}^{\prime})\cdot{\bf \sigma}||\leq |{\bf b}-{\bf b}^{\prime}|
\end{eqnarray}
and $2 \leq |{\bf b}+{\bf b}^{\prime}|+|{\bf b}-{\bf b}^{\prime}|\leq 2\sqrt{2}$, which is the upper bound in quantum mechanics and in fact realized by a singlet state. 
For any separable pure states, it is shown~\cite{werner} that 
\begin{eqnarray}
|\text{Tr}\rho_{separable}B|\leq 2.
\end{eqnarray}
See also eq.(4.19) below.
 
A possible local dispersion free representation for the quantum average by any pure state
$\rho=|\psi\rangle\langle \psi|$ is written as~\cite{bell2, chsh}
\begin{eqnarray}
\langle {\bf a}\cdot {\bf \sigma}\otimes {\bf b}\cdot {\bf \sigma}\rangle_{\psi}
=\int_{\Lambda} P(\lambda)a_{\psi}(\theta,\lambda)b_{\psi}(\varphi,\lambda)d\lambda
\end{eqnarray}
with dichotomic variables $a_{\psi}(\theta,\lambda)$ and $b_{\psi}(\varphi,\lambda)$ since four eigenvalues of ${\bf a}\cdot {\bf \sigma}\otimes {\bf b}\cdot {\bf \sigma}$ are
respectively $|{\bf a}||{\bf b}|$, $(-|{\bf a}|)|{\bf b}|$, $|{\bf a}|(-|{\bf b}|)$ and $(-|{\bf a}|)(-|{\bf b}|)$. 
Here we use $\lambda$ to represent a collection of possible hidden variables instead of $\omega$ in (2.2). Following Werner~\cite{werner},
we adopt this formula as the {\em definition} of the local hidden variables model of Bell and CHSH and examine the constraint imposed  on the weight $P(\lambda)$ by a minimum linearity requirement of quantum mechanics. This definition naturally agrees with Eq.(2) of Bell~\cite{bell2} and Eq.(3.5) of Clauser and Shimony~\cite{ch}. 
This model belongs to the so-called non-contextual model.
Contextuality is incorporated by using  $P_{\psi,\cal B}(\lambda)$ and 
$\Lambda_{\cal B}$ instead of $P(\lambda)$ and 
$\Lambda$ in (3.5), where the suffix ${\cal B}$ indicates the dependence of these quantities on the choice of ${\bf a}\cdot {\bf \sigma}\otimes {\bf 1}$, ${\bf 1}\otimes {\bf b}\cdot {\bf \sigma}$ and ${\bf a}\cdot {\bf \sigma}\otimes {\bf b}\cdot {\bf \sigma}$, and $\psi$ indicates the state dependence~\cite{beltrametti}. In contrast, the uniform  $P(\lambda)$  and $\Lambda$ for all the combinations of spin operators~\cite{bell2, chsh}  enforce the strict {\em locality}, namely, not only the factored form of $a_{\psi}(\theta,\lambda)b_{\psi}(\varphi,\lambda)$ but also no communication between the two parties through  $P(\lambda)$  and $\Lambda$. 
We mention the implications of contextuality later. The vectors ${\bf a}$ and ${\bf b}$ may be chosen to lie in the plane perpendicular to the relative momentum in the center of mass frame, and various azimuthal angles are defined in the plane around the relative momentum. 

We analyze the CHSH operator in the framework of the above local non-contextual hidden variables model in (3.5).
We first note that we can re-write the CHSH operator for {\em non-collinear} ${\bf b}$ and ${\bf b}^{\prime}$, which we study in the present paper, as 
\begin{eqnarray}
B&=&{\bf a}\cdot {\bf \sigma}\otimes ({\bf b}+{\bf b}^{\prime})\cdot {\bf \sigma} +{\bf a}^{\prime}\cdot{\bf \sigma}\otimes ({\bf b}-{\bf b}^{\prime})\cdot{\bf \sigma}\nonumber\\
&=& |{\bf b}+{\bf b}^{\prime}|[{\bf a}\cdot {\bf \sigma}\otimes \tilde{{\bf b}}\cdot {\bf \sigma}] +
|{\bf b}-{\bf b}^{\prime}|[{\bf a}^{\prime}\cdot{\bf \sigma}\otimes \tilde{{\bf b}}^{\prime}\cdot{\bf \sigma}]
\end{eqnarray}
by defining unit vectors
\begin{eqnarray}
\tilde{{\bf b}}=\frac{{\bf b}+{\bf b}^{\prime}}{|{\bf b}+{\bf b}^{\prime}|}, \ \  
\tilde{{\bf b}}^{\prime}=\frac{{\bf b}-{\bf b}^{\prime}}{|{\bf b}-{\bf b}^{\prime}|}, \ \ \tilde{{\bf b}}\cdot\tilde{{\bf b}}^{\prime}=0.   
\end{eqnarray}
By applying the local non-contextual hidden variables formula (3.5), we obtain
\begin{eqnarray}
\langle B \rangle_{\psi}
=\int P(\lambda)d\lambda [|{\bf b}+{\bf b}^{\prime}|a_{\psi}(\theta,\lambda)\tilde{b}_{\psi}(\phi,\lambda)+
|{\bf b}-{\bf b}^{\prime}|a_{\psi}(\theta^{\prime},\lambda)\tilde{b}^{\prime}_{\psi}(\phi^{\prime},\lambda)].\nonumber\\
\end{eqnarray}
Note that the assumption of non-contextuality is essential to write this relation.
By noting 
\begin{eqnarray}
&&-[ |{\bf b}+{\bf b}^{\prime}|+|{\bf b}-{\bf b}^{\prime}|]\nonumber\\
&& \leq [|{\bf b}+{\bf b}^{\prime}|a_{\psi}(\theta,\lambda)\tilde{b}_{\psi}(\phi,\lambda)+
|{\bf b}-{\bf b}^{\prime}|a_{\psi}(\theta^{\prime},\lambda)\tilde{b}^{\prime}_{\psi}(\phi^{\prime},\lambda)]\nonumber\\
&&\leq [ |{\bf b}+{\bf b}^{\prime}|+|{\bf b}-{\bf b}^{\prime}|] 
\end{eqnarray}
and the property for non-collinear ${\bf b}$ and ${\bf b}^{\prime}$
\begin{eqnarray}
2 < |{\bf b}+{\bf b}^{\prime}|+|{\bf b}-{\bf b}^{\prime}|\leq 2\sqrt{2}
\end{eqnarray}
we conclude
\begin{eqnarray}
|\langle B \rangle_{\psi}|\leq 2\sqrt{2}.
\end{eqnarray}

To achieve the upper or lower bound for (3.9), the existence of some domain in hidden variables space with 
$a_{\psi}(\theta,\lambda)\tilde{b}_{\psi}(\phi,\lambda)= a_{\psi}(\theta^{\prime},\lambda)\tilde{b}^{\prime}_{\psi}(\phi^{\prime},\lambda)=1$ or $a_{\psi}(\theta,\lambda)\tilde{b}_{\psi}(\phi,\lambda)=a_{\psi}(\theta^{\prime},\lambda)\tilde{b}^{\prime}_{\psi}(\phi^{\prime},\lambda)=-1$ is essential. If one assumes otherwise,  namely,
if $ a_{\psi}(\theta,\lambda)\tilde{b}_{\psi}(\phi,\lambda)=\pm 1 $ should always imply $a_{\psi}(\theta^{\prime},\lambda)\tilde{b}^{\prime}_{\psi}(\phi^{\prime},\lambda)=\mp1$ for any $\lambda$, respectively, the basic formula (3.5) would imply for a sum of two {\em non-commuting} operators 
\begin{eqnarray}
\langle {\bf a}\cdot {\bf \sigma}\otimes \tilde{{\bf b}}\cdot {\bf \sigma}\rangle_{\psi}
 +\langle {\bf a}^{\prime}\cdot{\bf \sigma}\otimes \tilde{{\bf b}}^{\prime}\cdot{\bf \sigma}\rangle_{\psi}=0.
\end{eqnarray}
This does not hold for  generic states $\psi$ if the formula (3.5) is sensible in the sense of quantum mechanics. This establishes the inequality (3.11). (The upper bound in (3.11) is achieved if one suitably chooses the weight $P(\lambda)$ with $\int P(\lambda)d\lambda=1$ which is peaked around the domain of $a_{\psi}(\theta,\lambda)\tilde{b}_{\psi}(\phi,\lambda)= a_{\psi}(\theta^{\prime},\lambda)\tilde{b}^{\prime}_{\psi}(\phi^{\prime},\lambda)=1$, for example.)

On the other hand, the conventional treatment~\cite{chsh} 
\begin{eqnarray}
\langle B\rangle_{\psi}&=& \langle {\bf a}\cdot {\bf \sigma}\otimes ({\bf b}+{\bf b}^{\prime})\cdot {\bf \sigma}\rangle
+\langle {\bf a}^{\prime}\cdot{\bf \sigma}\otimes ({\bf b}-{\bf b}^{\prime})\cdot{\bf \sigma}
\rangle\nonumber\\
&=&\int P(\lambda)d\lambda \{a_{\psi}(\theta,\lambda)[b_{\psi}(\varphi,\lambda)+b_{\psi}(\varphi^{\prime},\lambda)]\nonumber\\
&&+a_{\psi}(\theta^{\prime}, \lambda)[b_{\psi}(\varphi,\lambda)
-b_{\psi}(\varphi^{\prime},\lambda)]\}
\end{eqnarray}
uses the simultaneous dispersion free representations for non-commuting operators ${\bf a}\cdot {\bf \sigma}\otimes {\bf b}\cdot {\bf \sigma}$ and ${\bf a}\cdot {\bf \sigma}\otimes {\bf b}^{\prime}\cdot {\bf \sigma}$ in ${\bf a}\cdot {\bf \sigma}\otimes ({\bf b}+{\bf b}^{\prime})\cdot {\bf \sigma}$ (similarly in ${\bf a}^{\prime}\cdot{\bf \sigma}\otimes ({\bf b}-{\bf b}^{\prime})\cdot{\bf \sigma}
$ ) at each point of the hidden variables space.
By noting the relation for dichotomic variables
$a_{\psi}(\theta,\lambda)[b_{\psi}(\varphi,\lambda)+b_{\psi}(\varphi^{\prime},\lambda)]
+a_{\psi}(\theta^{\prime}, \lambda)[b_{\psi}(\varphi,\lambda)
-b_{\psi}(\varphi^{\prime},\lambda)]=\pm 2$,
we conclude that 
\begin{eqnarray}
|\langle B\rangle_{\psi} |\leq 2.
\end{eqnarray}
We note that the relation (3.13) is essentially the same as (1.1) if one replaces the integral $\int d\lambda$ by a summation $\sum_{\lambda}$.

The local non-contextual hidden variables model (3.5) of Bell and
CHSH thus predicts  (3.11) or (3.14), namely, $|\langle B\rangle_{\psi} |\leq 2\sqrt{2}
$ or $|\langle B\rangle_{\psi} |\leq 2$, for the identical quantum operator $B$ depending on the two different ways of evaluation.  We emphasize that the physical processes described by $\langle |{\bf b}+{\bf b}^{\prime}|[{\bf a}\cdot {\bf \sigma}\otimes\tilde{{\bf b}}\cdot{\bf \sigma}]\rangle_{\psi}$ and $\langle {\bf a}\cdot {\bf \sigma}\otimes {\bf b}\cdot{\bf \sigma}\rangle_{\psi}+\langle {\bf a}\cdot {\bf \sigma}\otimes {\bf b}^{\prime}\cdot{\bf \sigma}\rangle_{\psi}$ are quite different, but both of them are measurable and quantum mechanics tells that these two should always agree.  This discrepancy in the two predictions of the local hidden variables model means that the formula (3.5) does not satisfy the linearity relation 
\begin{eqnarray}
&&\langle {\bf a}\cdot {\bf \sigma}\otimes ({\bf b}\pm {\bf b}^{\prime})\cdot {\bf \sigma}\rangle
\nonumber\\
&&=\langle {\bf a}\cdot {\bf \sigma}\otimes {\bf b}\cdot {\bf \sigma}\rangle
 \pm \langle {\bf a}\cdot {\bf \sigma}\otimes {\bf b}^{\prime}\cdot {\bf \sigma}\rangle
\end{eqnarray}
for {\em non-collinear} ${\bf b}$ and ${\bf b}^{\prime}$ even after integration over hidden variables, in contrast to the case of (2.7) after integration. We emphasize that the linearity is a {\em local property} of quantum mechanics in contrast to entanglement.

On the basis of the above analysis, one may immediately conclude that the conventional CHSH inequality $|\langle B\rangle_{\psi} |\leq 2$ does not provide a reliable test of the local hidden variables model of Bell and CHSH. To be more precise,
one may conclude either (i) the local hidden variables model of Bell and CHSH in (3.5) contradicts quantum mechanics due to the failure of linearity (3.15) without referring to long-ranged EPR entanglement, or (ii) one needs to examine the consequences of the linearity condition (3.15) which renders the conventional CHSH inequality  $|\langle B\rangle_{\psi} |\leq 2$ as the unique prediction of the model (3.5). 
In the next section, we analyze the consequences of this linearity condition in detail.

In passing, we briefly comment on the original Bell's inequality~\cite{bell2}.
Bell's inequality deals with the operator ${\bf a}\cdot {\bf \sigma}\otimes({\bf b}-{\bf b}^{\prime})\cdot{\bf \sigma}$ and works on the same basis as (3.13). One then  starts with the numerical identity
\begin{eqnarray}
&&a_{\psi}(\theta, \lambda)[b_{\psi}(\varphi,\lambda)
-b_{\psi}(\varphi^{\prime},\lambda)]\nonumber\\
&&=a_{\psi}(\theta, \lambda)b_{\psi}(\varphi,\lambda)[1
-b_{\psi}(\varphi,\lambda)b_{\psi}(\varphi^{\prime},\lambda)]\nonumber\\
&&=\pm [1
-b_{\psi}(\varphi,\lambda)b_{\psi}(\varphi^{\prime},\lambda)],
\end{eqnarray}
combined with the condition $b_{\psi}(\varphi,\lambda)=-a_{\psi}(\varphi,\lambda)$ to account for a singlet state, namely,
\begin{eqnarray}
&&|\int P(\lambda)d\lambda a_{\psi}(\theta, \lambda)[b_{\psi}(\varphi,\lambda)
-b_{\psi}(\varphi^{\prime},\lambda)]|\nonumber\\
&&\leq
\int P(\lambda)d\lambda [1 +
a_{\psi}(\varphi,\lambda)b_{\psi}(\varphi^{\prime},\lambda)].
\end{eqnarray}
This relation, when converted to the quantum mechanical relation by means of (3.5), is known to contradict the predictions of quantum mechanics.

The purpose of Bell's original analysis is to show the inconsistency of the local hidden variables model with a singlet state in quantum mechanics and thus the singlet state is used as an essential element of the inequality, while the CHSH inequality in the local non-contextual hidden variables model is formulated without referring to the singlet state. This use of the {\em additional assumption} of the singlet state in Bell's formulation makes the analysis of the model (3.5) itself obscure. The inconsistency of the local hidden variables model in (3.5) with a quantum mechanical  singlet state is tested by the CHSH operator if one adopts the conventional inequality (3.14), which is based on the operation (3.13) just as the Bell's inequality.
 Bell's inequality and CHSH inequality are usually regarded to belong to the same class of inequalities, but they are rather  different when understood in the present manner. 

\section{Consequences of linearity}

\subsection{Linearity and contextuality}

We have shown that the local and non-contextual model (3.5) gives the two different predictions (3.11) and (3.14) for the identical quantum operator $B$ due to the failure of the linearity condition (3.15). Also,
Gleason's theorem~\cite{gleason} implies that we cannot define  fully 4-dimensional hidden variables models which are non-contextual. In view of this, we examine what happens if the linearity condition (3.15) is imposed on the local hidden variables model of Bell and CHSH and thereby resolving the above discrepancy. This amounts to specifying the weight factor $P(\lambda)$ in more detail and  inevitably leading to contextual models. We illustrate this phenomenon in the following.
\\

The linearity requirement (3.15)  imposes stringent consistency conditions on local dispersion free representations. For example, for {\em non-collinear} ${\bf b}$ and ${\bf b}^{\prime}$ which we study in the following, 
\begin{eqnarray}
&&\langle {\bf 1}\otimes ({\bf b}+{\bf b}^{\prime})\cdot {\bf \sigma}\rangle
\nonumber\\
&=&\int |{\bf b}+{\bf b}^{\prime}|\tilde{b}_{\psi}(\phi,\lambda)P(\lambda)d\lambda\\
&=&\int b_{\psi}(\varphi,\lambda)P(\lambda)d\lambda+\int b_{\psi}(\varphi^{\prime},\lambda)P(\lambda)d\lambda\nonumber
\end{eqnarray}
and 
\begin{eqnarray}
&&\langle {\bf a}\cdot {\bf \sigma}\otimes ({\bf b}+{\bf b}^{\prime})\cdot {\bf \sigma}\rangle
\nonumber\\
&=&\int a_{\psi}(\theta,\lambda)|{\bf b}+{\bf b}^{\prime}|\tilde{b}_{\psi}(\phi,\lambda)P(\lambda)d\lambda\\
&=&\int a_{\psi}(\theta,\lambda)[b_{\psi}(\varphi,\lambda)+ b_{\psi}(\varphi^{\prime},\lambda)]P(\lambda)d\lambda\nonumber
\end{eqnarray}
but the expressions local in $\lambda$ space are quite different as is shown by von Neumann's no-go argument, namely,
\begin{eqnarray}
|{\bf b}+{\bf b}^{\prime}|\tilde{b}_{\psi}(\phi,\lambda)\neq b_{\psi}(\varphi,\lambda)+ b_{\psi}(\varphi^{\prime},\lambda)
\end{eqnarray}
which is a general statement on the dispersion free representations of two non-commuting operators at {\em any point} in hidden variables space $\lambda$. See also (2.6). Note that the only integer allowed for 
$|{\bf b}+{\bf b}^{\prime}|$ is $|{\bf b}+{\bf b}^{\prime}|=1$ for non-collinear ${\bf b}$ and ${\bf b}^{\prime}$. The appearance of the same factor (4.3) in two different expressions (4.1) and (4.2) is a result of strict {\em
locality} assumption in (3.5). We have to satisfy two conditions (4.1) and (4.2) with the constraint (4.3). 

To make the mathematical analysis more transparent,  we parameterize the hidden variables and dichotomic variables as follows (which is suggested in the original paper of Bell~\cite{bell2}),
\begin{eqnarray}
&&P(\lambda)d\lambda=P(\lambda_{1},\lambda_{2})d\lambda_{1}d\lambda_{2},\nonumber\\
&&a_{\psi}(\theta,\lambda)= a_{\psi}(\theta,\lambda_{1}),\ \ \
b_{\psi}(\varphi,\lambda)=  b_{\psi}(\varphi,\lambda_{2}),\nonumber\\
&&\tilde{b}_{\psi}(\phi,\lambda)=  \tilde{b}_{\psi}(\phi,\lambda_{2}),\ \ \
\tilde{b}^{\prime}_{\psi}(\phi^{\prime},\lambda)= \tilde{b}^{\prime}_{\psi}(\phi^{\prime},\lambda_{2}). 
\end{eqnarray}
Namely, the $a$-system is parameterized by the hidden variables $\lambda_{1}$ and the $b$-system is parameterized by the hidden variables $\lambda_{2}$.  
We also define the "projection operator" by 
\begin{eqnarray}
A_{\psi}(\theta,\lambda_{1})=\frac{1}{2}[1 + a_{\psi}(\theta,\lambda_{1})]
\end{eqnarray}
which assumes $1$ or $0$. 

Then the conditions of the linearity (4.1)-(4.3) are summarized by 
\begin{eqnarray}
&&\int [|{\bf b}+{\bf b}^{\prime}|\tilde{b}_{\psi}(\phi,\lambda_{2})
-b_{\psi}(\varphi,\lambda_{2})-b_{\psi}(\varphi^{\prime},\lambda_{2})]\nonumber\\
&&\ \ \ \ \ \times P(\lambda_{1},\lambda_{2})d\lambda_{1}d\lambda_{2}=0,
\end{eqnarray}
\begin{eqnarray}
&&\int A_{\psi}(\theta,\lambda_{1})[|{\bf b}+{\bf b}^{\prime}|\tilde{b}_{\psi}(\phi,\lambda_{2})
-b_{\psi}(\varphi,\lambda_{2})-b_{\psi}(\varphi^{\prime},\lambda_{2})]\nonumber\\
&&\ \ \ \ \ \times P(\lambda_{1},\lambda_{2})d\lambda_{1}d\lambda_{2}=0,
\end{eqnarray}
and 
\begin{eqnarray}
&&|{\bf b}+{\bf b}^{\prime}|\tilde{b}_{\psi}(\phi,\lambda_{2})
-b_{\psi}(\varphi,\lambda_{2})-b_{\psi}(\varphi^{\prime},\lambda_{2})\neq 0,
\end{eqnarray}
for any pure state $\psi$ and for any  free parameters  $\theta$, $\varphi$ and $\varphi^{\prime}$, and in the case of (4.8) for any $\lambda_{2}$.

We would like to determine the possible structure of the weight function $P(\lambda_{1},\lambda_{2})$ from the conditions (4.6)-(4.8). From (4.6) and (4.7), we see that 
\begin{eqnarray}
&&P(\Lambda_{1}; \lambda_{2})=\int_{\Lambda_{1}}d\lambda_{1}P(\lambda_{1},\lambda_{2}), \nonumber\\
&&P(\psi, \theta; \lambda_{2})=\int_{\Lambda_{1}}d\lambda_{1}A_{\psi}(\theta,\lambda_{1})P(\lambda_{1},\lambda_{2}), \nonumber\\
&&\bar{P}(\psi, \theta; \lambda_{2})=\int_{\Lambda_{1}}d\lambda_{1}[1-A_{\psi}(\theta,\lambda_{1})]P(\lambda_{1},\lambda_{2}), 
\end{eqnarray}
define the weight factors of consistent (i.e., satisfying linearity) $d=2$ hidden variables models of the $b$-system, where $\Lambda_{1}$ is the entire space of the variables $\lambda_{1}$. The non-negative weight $P(\psi, \theta; \lambda_{2})$ receives the contribution from the domain with $A_{\psi}(\theta,\lambda_{1})=1$, and the non-negative  weight $\bar{P}(\psi, \theta; \lambda_{2})$ from its complement. For non-trivial hidden variables models, we have $P(\psi, \theta; \lambda_{2})\neq P(\Lambda_{1}; \lambda_{2})$ in general. In particular, (4.6) shows that $P(\Lambda_{1}; \lambda_{2})$ defines $d=2$ {\em non-contextual} hidden variables models.
Due to the symmetry between $a$-system and $b$-system, we can make a similar statement on $\lambda_{1}$ dependence.

If one assumes that the weights for the $d=2$ non-contextual hidden variables models are {\em uniquely} specified by the chosen dichotomic representations of $a_{\psi_{1}}(\theta,\lambda_{1})$ and $b_{\psi_{2}}(\varphi,\lambda_{2})$, respectively, which is the case of the known construction of hidden variables models in $d=2$~\cite{bell1, kochen}, one concludes from (4.9) a factored form of two systems
\begin{eqnarray}
P(\lambda_{1},\lambda_{2})=P_{1}(\lambda_{1})P_{2}(\lambda_{2})
\end{eqnarray}
where $P_{1}(\lambda_{1})$ stands for the weight of the $a$-system and 
$P_{2}(\lambda_{2})$ stands for the weight of the $b$-system in the sense of consistent non-contextual hidden variables models in $d=2$.  
For this choice in (4.10), $P(\psi, \theta; \lambda_{2})$ and $P(\Lambda_{1}; \lambda_{2})$ are equivalent as the weight for the $b$-system, and similarly for the $a$-system.  

We have so far analyzed the linearity constraint on the basis of the parameterization suggested by Bell in (4.4). A more general situation is handled if one denotes the hidden variables appearing in $a_{\psi}(\theta,\lambda)$ by $\lambda_{1}$, but keeping the dependence of other quantities on hidden variables general. In this case, the conditions in (4.6)-(4.8) are replaced by 
\begin{eqnarray}
&&\int [|{\bf b}+{\bf b}^{\prime}|\tilde{b}_{\psi}(\phi,\lambda_{1},\lambda_{2})
-b_{\psi}(\varphi,\lambda_{1},\lambda_{2})-b_{\psi}(\varphi^{\prime},\lambda_{1},\lambda_{2})],\nonumber\\
&&\ \ \ \ \ \times P(\lambda_{1},\lambda_{2})d\lambda_{1}d\lambda_{2}=0,\\
&&\int A_{\psi}(\theta,\lambda_{1})[|{\bf b}+{\bf b}^{\prime}|\tilde{b}_{\psi}(\phi,\lambda_{1},\lambda_{2})
-b_{\psi}(\varphi,\lambda_{1},\lambda_{2})-b_{\psi}(\varphi^{\prime},\lambda_{1},\lambda_{2})]\nonumber\\
&&\ \ \ \ \ \times P(\lambda_{1},\lambda_{2})d\lambda_{1}d\lambda_{2}=0,\\
&&|{\bf b}+{\bf b}^{\prime}|\tilde{b}_{\psi}(\phi,\lambda_{1},\lambda_{2})
-b_{\psi}(\varphi,\lambda_{1},\lambda_{2})-b_{\psi}(\varphi^{\prime},\lambda_{1},\lambda_{2})\neq 0.
\end{eqnarray}
If one wants to maintain the symmetry between 
$a$-system and $b$-system, one possibility is the absence of $\lambda_{2}$ dependence in (4.11)-(4.13), namely, the variables $\lambda_{1}$ describe the entire system uniformly. In this case, we have no integration over $\lambda_{2}$. It is then shown that (4.12) contradicts the relation (4.13) and the assumption of non-contextuality, and thus this case is excluded. 
This is seen by considering a specific separable state
\begin{eqnarray}
\psi=\psi_{1}({\bf s}_{1})\psi_{2}({\bf s}_{2})
\end{eqnarray}
by assuming non-contextuality, namely, the same $P(\lambda_{1})
$ is valid for any state $\psi$  and any operator ${\bf a}\cdot {\bf \sigma}\otimes {\bf b}\cdot {\bf \sigma}$. For the specific state in (4.14),  one can make  $A_{\psi_{1}}(\theta,\lambda_{1})=1$ for any $\lambda_{1}\in \Lambda_{1}$ by choosing the orientation of the spin perfectly correlated with the wave function ${\bf a}={\bf s}_{1}$. One may then gradually change $\theta$ such that one eventually arrives at the anti-parallel configuration with $A_{\psi_{1}}(\theta,\lambda_{1})=0$ for any $\lambda_{1}\in \Lambda_{1}$. By this procedure, we can scan the entire spectrum (i.e., $\lambda_{1}$ dependence) of $P(\lambda_{1})$ by keeping the condition in (4.12) in tact for the separable state (4.14). Namely,
the combination in (4.12) without $\lambda_{2}$ dependence 
\begin{eqnarray}
A_{\psi_{1}}(\theta,\lambda_{1})P(\lambda_{1})
\end{eqnarray}
measures the $\lambda_{1}$ dependence of $P(\lambda_{1})$ in the domain with $A_{\psi_{1}}(\theta,\lambda_{1})=1$, and this area changes from the entire domain $\Lambda_{1}$ to the null in the above process, by keeping the condition (4.12) in tact. This gives the relation
\begin{eqnarray}
&&[|{\bf b}+{\bf b}^{\prime}|\tilde{b}_{\psi_{2}}(\phi,\lambda_{1})
-b_{\psi_{2}}(\varphi,\lambda_{1})-b_{\psi_{2}}(\varphi^{\prime},\lambda_{1})=0,
\end{eqnarray}
at least for a point $\lambda_{1}\in \Lambda_{1}$. But this contradicts (4.13) with $\psi=\psi_{2}$ without $\lambda_{2}$ dependence for non-collinear  ${\bf b}$ and ${\bf b}^{\prime}$.
The other symmetric possibility is reduced to the Bell's parameterization in (4.4) we have already analyzed. It is interesting that  the Bell's parameterization in (4.4) is the only consistent parameterization if one imposes non-contextuality and linearity requirements. (If one accepts Gleason's theorem, the 
non-contextual $P(\lambda_{1})$ in $d=4$ is excluded, but $P(\lambda_{1},\lambda_{2})=P_{1}(\lambda_{1})P_{2}(\lambda_{2})$ in (4.10) is basically contextual in the sense of $d=4$ and thus allowed.)

The basic formula of Bell and CHSH in (3.5) is now written by using (4.10) as 
\begin{eqnarray}
\langle {\bf a}\cdot {\bf \sigma}\otimes {\bf b}\cdot {\bf \sigma}\rangle_{\psi}
=\int_{\Lambda_{1}} P_{1}(\lambda_{1})a_{\psi}(\theta,\lambda_{1})d\lambda_{1}\int_{\Lambda_{2}}P_{2}(\lambda_{2})b_{\psi}(\varphi,\lambda_{2})d\lambda_{2}
\end{eqnarray}
which was originally supposed to be valid for the general state
$\rho=|\psi\rangle\langle \psi|$
where $|\psi\rangle$ is a 4-dimensional pure state,
but it is actually valid only for the pure separable state, 
\begin{eqnarray}
\rho=|\psi_{1}\rangle\langle\psi_{1}|\otimes|\psi_{2}\rangle\langle\psi_{2}|
\end{eqnarray}
due to the quantum mechanical linearity condition (3.15). In terms of the language of hidden variables models on the right-hand side of (4.17), the formula is based on a factored product of two
consistent non-contextual $d=2$ hidden variables models such as defined in~\cite{bell1,kochen}.
This contextual hidden variables model in the sense of $P_{\psi}(\lambda_{1},\lambda_{2})$ in $d=4$ does not contradict Gleason's theorem~\cite{gleason}.

The local hidden variables model (3.5) is supposed to describe the purely local properties of quantum mechanics, but one finds that it describes a certain aspect of (long-ranged) entanglement if one uses a suitable operator such as (3.1) and compares its two alternative ways of evaluation in (3.11) and (3.14). The linearity requirement eliminates this remnant of entanglement in the formula (3.5) and leads to the manifestly separable quantum states as the only allowed states.

\subsection{Separable quantum states}

We have shown that the local non-contextual model (3.5) of Bell and CHSH, when the linearity condition is imposed, is reduced to a factored product of two consistent non-contextual hidden variables models
in $d=2$. As for our starting problem, namely, two different expressions (3.11) and (3.14) for the same CHSH operator $B$, it is resolved  in the case of the separable model in (4.17) since  the linearity condition (3.15) is satisfied by a factored form of two non-contextual $d=2$ hidden variables models. This resolution by a factored product may appear to be  rather trivial, but the construction of $d=2$ non-contextual models which satisfy the linearity condition is highly non-trivial~\cite{bell1,kochen}. Both of (3.11) and (3.14) thus give the same expression
\begin{eqnarray}
\langle B\rangle&=&\int P_{1}(\lambda_{1})P_{2}(\lambda_{2})d\lambda_{1} d\lambda_{2}\nonumber\\
&&\times\{a_{\psi_{1}}(\theta,\lambda_{1})[b_{\psi_{2}}(\varphi,\lambda_{2})+b_{\psi_{2}}(\varphi^{\prime},\lambda_{2})]\nonumber\\
&&+a_{\psi_{1}}(\theta^{\prime},\lambda_{1})[b_{\psi_{2}}(\varphi,\lambda_{2})-b_{\psi_{2}}(\varphi^{\prime},\lambda_{2})]\}.
\end{eqnarray}
If one notes, 
$a_{\psi_{1}}(\theta,\lambda_{1})[b_{\psi_{2}}(\varphi,\lambda_{2})+b_{\psi_{2}}(\varphi^{\prime},\lambda_{2})]\nonumber\\
+a_{\psi_{1}}(\theta^{\prime},\lambda_{1})[b_{\psi_{2}}(\varphi,\lambda_{2})-b_{\psi_{2}}(\varphi^{\prime},\lambda_{2})]=\pm 2$,
one recovers the ordinary CHSH inequality $|\langle B\rangle| \leq 2$, and it also gives a proof of the quantum mechanical inequality for a separable state in (3.4) by means of  consistent $d=2$ hidden variables models such as in (2,2). Werner showed that the conventional CHSH inequality gives a necessary and sufficient separability criterion of pure states in the framework of quantum mechanical states~\cite{werner}. We arrived at the separable pure states in the framework of $d=4$ local hidden variables models with linearity condition added.   

Experimental findings, although in the context of two photon correlation~\cite{aspect}, indicate the violation of CHSH inequality in the sense $|\langle B\rangle| \leq 2$
implied by separable states in $d=4$; they definitely show that the full contents of quantum mechanics even for a far-apart system cannot be described by separable states only. We also emphasize that the experiment does not exclude consistent  $d=2$ non-contextual hidden variables models themselves.  See, however, the recent analysis~\cite{fujikawa}.

We next mention an interesting application of CHSH inequality to quantum cryptography by Ekert~\cite{ekert}, which is based on the mixed separable states 
\begin{eqnarray}
\rho=\int d{\bf n}_{a}d{\bf n}_{b}w({\bf n}_{a},{\bf n}_{b})\rho({\bf n}_{a})\otimes\rho({\bf n}_{b}),
\end{eqnarray} 
and satisfies  the relation  $-2\leq{\rm Tr}[\rho B]\leq2$.
This formula is a purely quantum mechanical one with no reference to  dispersion free representations.
If one considers the pure separable state $\psi=\psi_{1}({\bf n}_{a})\psi_{2}({\bf n}_{b})$  and writes the corresponding density matrix as $\rho({\bf n}_{a})\otimes\rho({\bf n}_{b})$, we have the defining relation
\begin{eqnarray}
\langle {\bf a}\cdot {\bf \sigma}\otimes {\bf b}\cdot {\bf \sigma}\rangle_{\psi}&=&\int_{\Lambda_{1}} P_{1}(\lambda_{1})a_{\psi_{1}}(\theta,\lambda_{1})d\lambda_{1}\int_{\Lambda_{2}}P_{2}(\lambda_{2})b_{\psi_{2}}(\varphi,\lambda_{2})d\lambda_{2}\nonumber\\
&=&{\rm Tr}\{[{\bf a}\cdot {\bf \sigma}\otimes {\bf b}\cdot{\bf \sigma}][\rho({\bf n}_{a})\otimes\rho({\bf n}_{b})]\}\nonumber\\
&=&({\bf a}\cdot {\bf n}_{a})({\bf b}\cdot {\bf n}_{b}),
\end{eqnarray}
while the mixed separable state (4.20) gives
\begin{eqnarray}
\langle {\bf a}\cdot {\bf \sigma}\otimes {\bf b}\cdot {\bf \sigma}\rangle_{\rho}
&=&\int d{\bf n}_{a}d{\bf n}_{b}w({\bf n}_{a},{\bf n}_{b})({\bf a}\cdot {\bf n}_{a})({\bf b}\cdot {\bf n}_{b}).
\end{eqnarray}
It is important to recognize that classical vector quantities ${\bf n}_{a}$ and ${\bf n}_{b}$ have no direct connection with hidden variables such as $\lambda_{1}$ and $\lambda_{2}$.

\section{Discussion and conclusion}

We have shown that the formula of Bell and CHSH in (3.5) does not define a consistent local non-contextual hidden variables model, as is evidenced by the fact that it leads to two different predictions for the quantum CHSH operator $B$ depending on the two different ways of evaluation, and thus the conventional CHSH inequality  $|\langle B\rangle |\leq 2$ does not provide a reliable test of the local non-contextual hidden variables model. The formula (3.5) defines certain correlations but they have little to do with quantum mechanics even with respect to local properties such as linearity. The supposedly non-contextual formula (3.5) in $d=4$ simply does not exist. One might still argue that it is meaningful to compare the non-quantum mechanical predictions of (3.5) with quantum mechanical experiments. This appears to be a rather common view, and then the prediction $|\langle B\rangle |\leq 2\sqrt{2}$ in (3.11) agrees with experiments but the prediction $|\langle B\rangle |\leq 2$ in (3.14) disagrees with experiments~\cite{aspect}.   

To make the model (3.5) consistent, one needs to impose the minimum linearity requirement on the model, which converts the non-contextual model to contextual one. We illustrated this phenomenon by showing that the formula (3.5) of Bell and CHSH is reduced to  a factored product of two $d=2$ non-contextual hidden variables models, which defines 
quantum mechanical separable states and satisfies the conventional CHSH inequality  $|\langle B\rangle |\leq 2$. 

In conclusion, combined with the fact that  the conventional CHSH inequality gives the necessary and sufficient criterion of separability for pure quantum mechanical states~\cite{werner}, it is our  opinion  that we should interpret the experimental refutation~\cite{aspect} of the conventional CHSH inequality as a proof that  the full contents of quantum mechanics even for a far-apart system cannot be described by separable quantum mechanical states only, instead of referring to the ill-defined local non-contextual hidden variables model (3.5) in $d=4$. 
 
It may also be worth adding that the very recent analysis~\cite{fujikawa} of conditional measurements excludes any non-contextual hidden variables models including those in $d=2$ such as the ones of Bell~\cite{bell1} and Kochen-Specker~\cite{kochen}, and thus no viable non-contextual hidden variables models whatsoever: There exist no known viable non-contextual hidden variables models that the convetional CHSH inequality might test.


\begin{thebibliography}{99}
\bibitem{bell2}
J. S. Bell, Physics {\bf 1}, 195 (1965).
\bibitem{chsh}
J. F. Clauser, M. A. Horne, A. Shimony and R. A. Holt, Phys. Rev. Lett. {\bf 23},  888 (1969).
\bibitem{epr}
A. Einstein, B. Podolsky and N. Rosen, Phys. Rev. {\bf 47}, 777 (1935).
\bibitem{peres}
A. Peres, {\em Quantum Theory: Concepts and Methods }, (Kluwer Academic Pub., 1995).
\bibitem{gleason}
A. M. Gleason, J. Math. Mech. {\bf 6}, 885 (1957).
\bibitem{bell1}
J. S. Bell, Rev. Mod. Phys. {\bf 38}, 447 (1966).
\bibitem{kochen}
S. Kochen and E. P. Specker, J. Math. Mech. {\bf 17}, 59 (1967).
\bibitem{ch}
The early history of Bell and CHSH inequalities is well reviewed
in, J.F. Clauser and A. Shimony, Rep. Prog. Phys. {\bf 41}, 1881 (1978).
\bibitem{cirel'son}
B.S. Cirel'son, Lett. Math. Phys. {\bf 4} (1980) 93.
\bibitem{werner}
R.F. Werner, Phys. Rev. A{\bf 40}, 4277 (1989).
\bibitem{beltrametti}
E. G. Beltrametti and G. Gassinelli, {\em The Logic of Quantum 
Mechanics}, (Addison-Wesley Pub., 1981).
\bibitem{fujikawa}
If one postulates that any physical quantity should have a unique expression in hidden variables space just as any quantum mechanical quantity has a unique space-time dependence, all the non-contextual hidden variables models including those in $d=2$, such as the models of Bell~\cite{bell1} and Kochen-Specker~\cite{kochen}, are excluded.    
 See \\
K. Fujikawa, Phys. Rev. A85 (2012) 012114; arXiv:1201.4421[quant-ph].
\bibitem{neumann}
J. von Neumann, {\em Mathematical Foundations of Quantum Mechanics}
(Princeton Univ. Press, 1955).
\bibitem{mermin}
N.D. Mermin, Phys. Rev. Lett. {\bf 65}, 3373 (1990).
\bibitem{aspect}
A. Aspect, J. Dalibard and G. Roger, Phys. Rev. Lett. {\bf 49},
1804 (1982).
\bibitem{ekert}
A.K. Ekert, Phys. Rev. Lett. {\bf 67}, 661 (1991).
\end{thebibliography}
\end{document}